**Adam POPOWICZ**

Silesian University of Technology, Institute of Electronics

# Optical identification of crystal defects in CCD matrix

***Streszczenie.*** *Celem artykułu jest zaprezentowanie oryginalnego pomysłu identyfikacji typów defektów struktury krystalicznej czujników światła jakimi są matryce CCD. Procedura jest nieskomplikowana i możliwa do przeprowadzenia bez specjalistycznej i drogiej aparatury. Metoda ta umożliwia rozróżnienie defektów na: punktowe oraz defekty przestrzenne – dyslokacje. Podczas badań wykazano również, iż typ defektu wpływa na zachowanie generacji prądu ciemnego podczas rejestracji światła.* (**Identyfikacja defektów struktury krystalicznej matryce CCD**)

***Abstract****. An original idea of semiconductor defects identification in CCD matrix was presented in the article. The procedure is simple and easy to execute because of no need for special and expensive equipment. The method classifies defects into two groups: the point defects and the spatial defects (dislocations). During the experiments it was proven that the type of defect affects the behavior of the dark current generation during the light gathering .*

**Słowa kluczowe**: matryca CCD, defekty w półprzewodniku, prąd ciemny

## Introduction

Nowadays CCD matrices are mainly used for imaging. Although they have been driven out of the consumer market by CMOS sensors [1], they seem to have firm position in scientific applications [2]. They were initially expected for use in analog systems [3] but soon their light registration capabilities were discovered. A special place of CCD usage is an observational astronomy where long-term exposure can record low-light cosmic structures [4, 5]. The CCD systems completely replaced the previously used photographic film due to the incomparably greater light sensitivity and the accuracy. They are often sent into the space providing information of key importance in understanding the universe.

On the other hand CCD sensor can be used for the analysis of various phenomena in semiconductors such as the two-photon absorption [6], the analysis of unstable defects [7], the influence of cosmic rays on the crystal structure of silicon [8, 9] or the activation energy of the thermal generation processes [10, 11]. CCD arrays are also used for the detection of the elementary particles [12].

## Crystal structure defects in CCD

Among the above-mentioned, the crystal defects are especially investigated in the article. A large number of pixels in CCD matrix (reaching several millions) is doubtless an advantage because many measurement can be made. Basically there are two types of defects (Fig. 1) : the point defects (foreign atoms, atomic gap – vacancy) and the spatial defects (dislocations) [13, 14, 15].

The point defects can occur during CCD production (the impurities in technology processes) but also can be the result of the high-energy particles striking the matrix and causing a vacancy or an atomic displacement as a result of a collision. The spatial defects can occur only during the fabrication process for example in Czochralski crystals growth method [16].

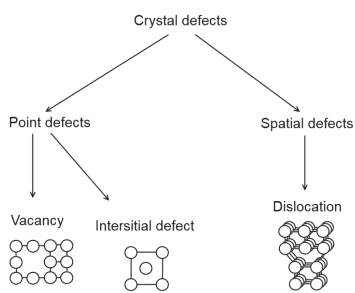

Fig. 1. Crystal defects classification

The widely used method for crystal defects analysis is the capacitance spectroscopy DLTS (Deep Level Transient Spectroscopy) [17]. The method allows analyzing defects in the p-n junction during depletion by appropriate bias voltage waveforms. The trap states in silicon associated with the defects are loaded and unloaded giving changes in the p-n capacitance which yield information about the defects (the thermal ionization energy, the capture cross section of electrons, the concentration of defects).

However this method is almost impossible to use in context of defects in the CDD pixels because of a lack of connections to each pixel individually and a number of necessary measurements reaching millions. It was therefore decided to create a simple method to classify the types of defects dominating in the CCD.

## Thermal generation

One of the most important effects of presence of defects in CCD pixel is the dark current causing the accumulation of additional charge not related to the photoelectric effect [11]. This process is one of the additive noise sources which must be corrected in order to obtain a precise measurement of the light [18]. This is especially important for imaging in astronomy, where long exposures (reaching often hour in astronomical spectroscopy) contribute to the accumulation of a large thermal charge. To enable such a registration, astronomical cameras are equipped with efficient cooling systems, which reduce the rate of thermal generation.

Research on the thermal generation of point defects leads to separation of groups associated with the presence of foreign atoms (carbon, nickel, cobalt, gold) and vacancies. The measurements are based on comparison of the generation rate at various temperatures. An activation energy of the process is found from the Arrhenius law [10]:

$$(1) \quad v = v_0 exp\left(\frac{-E_A}{kT}\right)$$

where: $v$ – velocity of the process, $v_0$ – constant, $E_A$ – activation energy [eV], $k$ – Boltzmann constant [eV/K], $T$ – temperature [K].

Obtained activation energies are specific to particular types of point defects. However such thermal analysis is possible if the individual pixels contain only one point defect. In case of more defects ($N > 1$) Arrhenius equation becomes:

$$(2) \quad v = \sum_{i=1}^{N}\left(v_i exp\left(\frac{-E_{A_i}}{kT}\right)\right)$$

In such situation results are difficult to analyze and it is not possible to distinguish the presence of dislocations from the presence of a group of different point defects. Moreover, the dislocations may be smaller or larger, so that the thermal generation rate of point defects is very similar to the generation rate of dislocations.

**Nonlinearities of dark current due to dislocations**

The idea of dislocation detection arose during the research on dark current exposure time nonlinearities [19, 20]. These works and also the articles [21, 22] have shown that specific defects seating on the edge of the depletion area in a pixel can cause a gradual decrease in the rate of thermal generation due to the shrinking of the depleted area. The proposed correction method based on these observations was successively verified on the astronomical pictures taken with the astronomical camera SBIG STL 11000M [20]. In the further research, using new matrix KAF8300 in ATIK383L+ camera, the author noticed a group of pixels behaving slightly differently. Nonlinearities were smoother unlike in the previous camera. Both the methods: proposed in [20] and the standard correction using dark frames [4, 2, 18] were not able to estimate the thermal charge correctly. The correction of uniformly illuminated picture with exposure time of 10 minutes was tested. After applying the correction, every nonlinear pixel's brightness was compared to the averaged brightness of the neighboring pixels. Correction method [20] underestimated the thermal charge (the brightness in a corrected pixel was higher than in surroundings pixels) and standard dark frame subtraction overestimated thermal charge (the brightness in a corrected pixel was lower than in surroundings).This group of pixels has also been noticed in the work [21]. The authors concluded that this group (in the article – group 3) is poorly understood and they could not explain the source of such nonlinearities.

The problem of the smooth nonlinearities of the dark current can be solved by the analysis of Barrier Recombination Model (Polish Academy of Sciences) extensively described and confirmed by the observations of changes in the photoconductivity [13]. The proposed mechanism of thermal generation concerns spatial defects (dislocations). It turned out that the dislocation interacts electrically with the crystal structure creating a potential barrier around itself and gathering positive charges (holes). This can cause a gradual change in the conditions of thermal generation in CCD's pixels so that it becomes increasingly slower. This is due to the recombination of the trapped electrons with the holes collected within dislocation potential barrier. The main recombination processes are shown in the Figure 2.

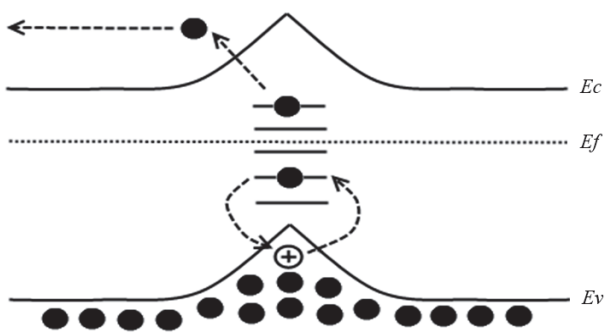

Fig. 2. The recombination processes and the electronic band structure of a dislocation. *Ec, Ef* and *Ev* are respectively: the conduction energy level, the Fermi energy level and the valence band energy level

One can see how the electrons stored in the trap states are either moving to the conduction band and automatically attracted to the pixel potential well, or recombining with the holes (if there was no energy level bend, the holes would flow to the neutral ground preventing such recombination). The main reason of growing number of the holes is the electron transfer to the conduction band. It is because every trapped electron leaves a hole in the valence band and when it is taken to the potential well, instated of recombining, there are excess holes in the dislocation cave. Finally, the generation gradually decreases because of the increase in hole recombination.

As in the case of point defects present on the edge of the depletion, generation will decrease due to light illumination; however the reason is not shrinking of the depleted area. As a result of the photoelectric effect, both: the electrons in conduction band and the holes in valence band appear. Only small part of the hole stream is absorbed by the dislocation barrier. If the whole stream of the holes was captured in dislocation cave, there would be a direct connection between the number of electrons in potential well and the level of thermal generation, as it was presented in previous work [20]. Actually, the number of captured holes is an unknown fraction of total number of holes in the stream so the decrease of the thermal generation rate is lower than predicted by [20].

From the presented analysis it can be concluded that the difference in the crystal defects lies in the dark current rate response to illumination level. For the point defects this relationship is independent of the pixel charge source (either thermal or photoelectric) [19, 20]:

(3) $$g_{dark} = \frac{dQ_d}{dt} = h(Q_c)$$

where: $Q_d$ – thermally generated charge in pixel well, $t$ – time, $Q_c$ – whole charge in pixel well.

In the case of dislocations the situation is much more complicated. Full description of the phenomenon would require knowledge of some dislocation properties such as its position and the arrangement in depletion area. However, from the point of view of this article, the most important is the fact, that the thermal generation rate as a function of the charge collected in the pixel's well is minimal when the charge comes from the thermal generation. It increases as a light induced charge participates in whole collected charge:

(4) $$g_{dark} = \frac{dQ_d}{dt} = h(Q_d, Q_c)$$

$$h(Q_d, Q_c) \to \min \Leftrightarrow Q_d \to Q_c$$

**Initial sensor lightening**

For the identification process differentiating sources of the dark current, the time dependencies of dark current where used [19, 20, 21]. The exposure times for taken dark frames were as follows: 0s, (the offset frame – bias), 10s, 20s, 40s, 80s, 160s, 320s ,640s, 1000s, 2000s, 3000s. Two cooled astronomical cameras were used: ATIK 11000M with KAI11002 interline CCD sensor and ATIK 383L+ with KAF8300 full-frame CCD. The dark current exposure time dependencies for every pixel were prepared. After the proper transformation [20], the relationships between the rate of generation and total thermal charge were computed and called $h(I_d)$. Thus, in case of linear time dependency $h$ should be constant, but for nonlinear dependency it is more complicated, showing a decrease in the dark current rate.

The examples of the dark current time dependencies and appropriate *h* relationships can be found in the Figure 3.

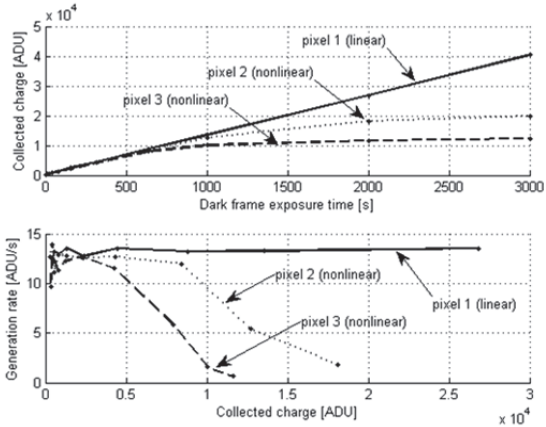

Fig. 3. Sample dark current exposure time dependencies and calculated *h* relationships for exemplar pixels (one linear and two nonlinear) in KAI 11002M CCD

The method of separation of thermal generation sources is based on comparison of *h* function determined from dark current exposure time dependencies (as in figure above) with the function *h'* obtained by pre-loading pixel with initial charge (light flash) and observing the following dark current rate. The measurements were made in the following manner: first – initial short light pulse with controlled intensity was provided (2 seconds – the light registration part), and then, without interrupting the exposure, light was turned off so that only thermal generation appeared (following 20 seconds – the dark registration part). Flat-field light panel (AURORA) was used as light source. Amount of the thermal charge in selected "hot" pixels (ie. having over average generation rate) was determined as the difference between the chosen pixel's charge and an average charge of neighboring pixels (the pixels adjacent to the chosen pixel). Such a method for dark current estimation is justified by the fact that the amount of dark current in pixels without defects was negligible and changes in light flux were virtually impossible for such a small area (15μm x 15μm for KAF8300 and 27μm x 27μm for KAI 11002M).

Finally, from each measurement, the pair of results was obtained for every pixel: the average initial charge in well (charge in neighboring pixels) and the rate of thermal generation (the amount of computed thermal charge divided by 20 seconds). The measurement was repeated 100 times changing initial light flux to cover the range of possible initial charge in well from emptiness to ¾ capacity. Therefore 100 points of function *h'* was obtained for every hot pixel.

**Data analysis**

The main result of the above-described experiments was the estimation of function *h'*. The next step was the comparison of the results of *h'* function obtained during the exposure with the results *h* function obtained by analyzing the time dependence of dark current. Resultant points of both functions were linearly approximated. For exemplary pixels with nonlinear exposure time dependencies of dark current a full set of results is presented in Figure 4 and Figure 5.

For point defect, the points from both, function *h* and *h'* seem to overlap. Therefore, there will be seen an equality between both approximations (as in Figure 4). However, the dislocation rate of the thermal generation should be higher in the experiments with the flash then in the calculation derived from the time dependencies (as in Figure 5).

To compare the obtained approximation the author decided to create a coefficient called "generation rate field".

It was chosen as an area under linear approximation. Such a coefficient was used because it presents averaged thermal generation rate. Point defects should have similar generation field coefficients for both function *h* and *h'* while for dislocation they should differ.

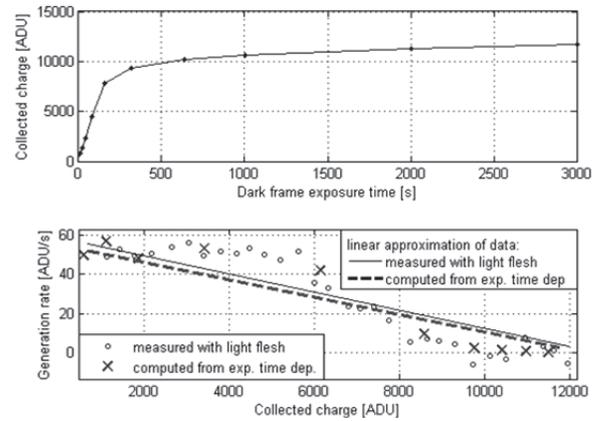

Fig. 4. Full set of results for typical point defect (KAI11002M)

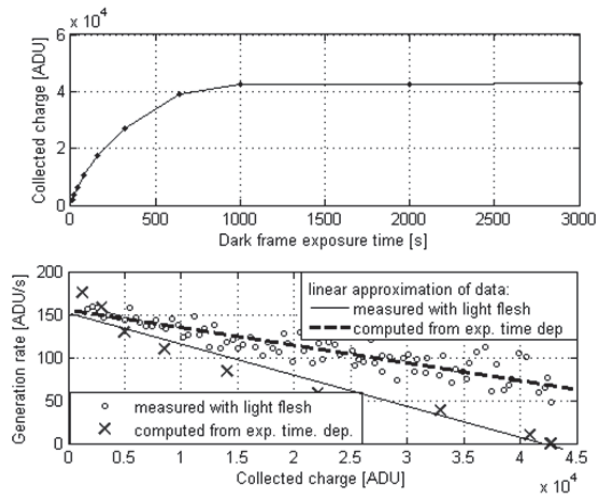

Fig. 5. Full set of results for typical dislocation defect (KAF8300)

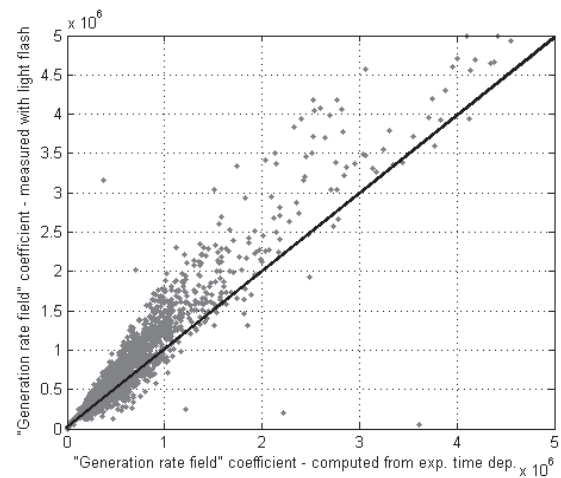

Fig. 6. Generation area coefficients for KAF8300 CCD

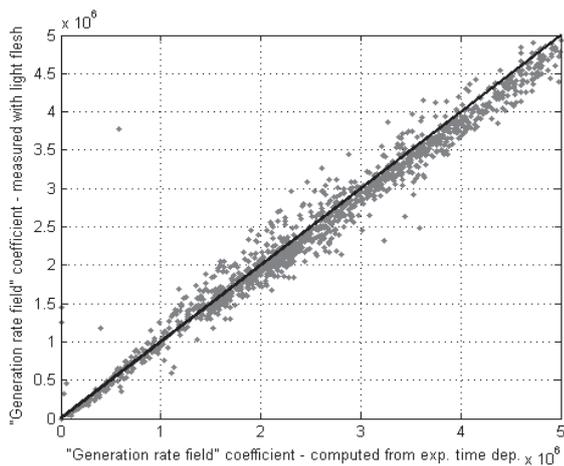

Fig. 7. Generation field coefficients for KAF8300 CCD

The Figures 6 and 7 confirmed the differences in the sources of thermal generation between the two matrices used. In the KAI11002M points arranged in a straight line around the *y=x* relationship, which suggests the presence of the point defects. It is in line with the results obtained in previous studies [19, 20]. The results of KAF8300 analysis look different. There is no clear linear relationship between the generation filed coefficients. Most of the generation field coefficients of the *h'* approximation are higher than of the *h* approximation. It confirms previous predictions of higher thermal generation from dislocations when the collected charge is induced by the light then when it is thermally induced.

**Conclusions**

The article presents a new approach to the analysis of the thermal generation sources in CCDs. The proposed methodology allows to distinguish causes of the dark current. It evaluates general tendencies in analyzed CCD matrix. Two types of the crystal defects were considered: the point defects (like intrinsic and interstitial atoms or the vacancies) and the spatial defects – dislocations. The method is easy implementable and does not require specialized equipment (like in DLTS method). It showed noticeable differences in both analyzed CCDs: Kodak KAF8300 full-frame CCD, and Kodak KAI11000M interline CCD. It seems that the dislocations as the source of the CCD dark current has not been adequately studied yet and this article can provide basis for further research.

***Author***: *M. Sc. Adam Popowicz, Silesian University of Technology, Institute of Electronics, Akademicka 16, 44-100 Gliwice, E-mail: apopowicz@polsl.pl;*